\documentclass[10pt,twocolumn]{acmish}

\usepackage{tabularx}
\usepackage{color,colortbl}
\usepackage[none]{hyphenat}

\author{
Steven Y. Ko\\
{\small Computer Science and Engineering}\\
{\small University at Buffalo}\\
{\small \textsf{stevko@buffalo.edu}}
\and
Lauren Sassoubre\\
{\small Civil, Structural, and Environmental Engineering}\\
{\small University at Buffalo}\\
{\small \textsf{lsassoub@buffalo.edu}}
\and
Jaroslaw Zola\footnotemark[0]\\
{\small Computer Science and Engineering}\\
{\small Biomedical Informatics}\\
{\small University at Buffalo}\\
{\small \textsf{jzola@buffalo.edu}}
}
\footnotetext[0]{To whom correspondence should be addressed.}

\title{Applications and Challenges of Real-time\\ Mobile DNA Analysis}

\begin{document}
\maketitle

\begin{abstract}
The DNA sequencing is the process of identifying the exact order of nucleotides within a given DNA molecule. The new portable and relatively inexpensive DNA sequencers, such as Oxford Nanopore MinION, have the potential to move DNA sequencing outside of laboratory, leading to faster and more accessible DNA-based diagnostics. However, portable DNA sequencing and analysis are challenging for mobile systems, owing to high data throughputs and computationally intensive processing performed in environments with unreliable connectivity and power.

In this paper, we provide an analysis of the challenges that mobile systems and mobile computing must address to maximize the potential of portable DNA sequencing, and {\it in situ} DNA analysis. We explain the DNA sequencing process and highlight the main differences between traditional and portable DNA sequencing in the context of the actual and envisioned applications. We look at the identified challenges from the perspective of both algorithms and systems design, showing the need for careful co-design.
\end{abstract}

\section{Introduction}

DNA, a polymer made from four basic nucleotides (abbreviated by A, C, G, T), is the main carrier of the genetic information. The DNA sequencing is the process in which this information is extracted by converting physical DNA molecules into a signal that describes the exact order and type of the constituent nucleotides. The ability to sequence DNA has revolutionized molecular biology, biomedicine and life sciences in general. Among many applications, some of which we review in Section~\ref{sec:applications}, it is recognized as a critical method for diagnosing and improving human health (e.g. dissecting genetic mechanisms of cancer~\cite{Stratton2009}), identifying pathogens and protecting public health (e.g. detecting and tracking spread of infectious diseases~\cite{Gardy2017}) or understanding our environment (e.g. impact of microorganisms on water, air~and~soil~\cite{Quince2017}).

\begin{figure}[b]
\centering
\scalebox{0.9}{
  \includegraphics{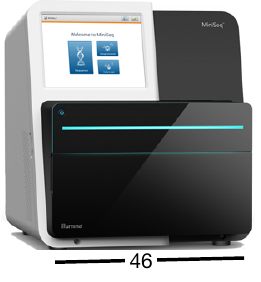}\hspace{0.2in}\includegraphics{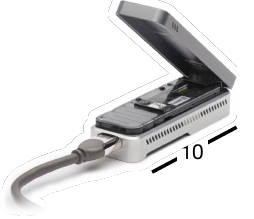}\hspace{0.05in}\includegraphics{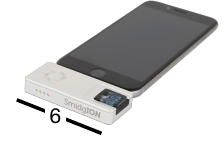}
}
\caption{MiniSeq DNA sequencer from Illumina (left), vs. portable MinION (middle) and forthcoming SmidgION (right). Pictures not to scale. The approximate size in centimeters is marked for reference. The current price of MinION is \$1,000 compared to \$50K for an entry-level benchtop sequencer.}\label{fig:ont}
\end{figure}

The end-to-end DNA sequencing and analysis involves a combination of laboratory and bioinformatics steps (see Section~\ref{sec:sequencing}). In~a~traditional setup, these steps are performed at massive scales by highly trained personnel using expensive benchtop DNA sequencers and supporting computational servers. Consequently, the process has been confined to high-end laboratories with financial resources and skilled personnel, limiting the access and extending the time it takes from sample collection to results.

The recently introduced portable DNA sequencers, specifically Oxford Nanopore Technology (ONT) MinION~\cite{ONT} (see Fig.~\ref{fig:ont}), are changing this situation. Compared to the traditional DNA sequencers, these devices are relatively inexpensive and truly mobile: smaller than a cell phone, USB powered, and designed to be easily operable ``in the field.'' Moreover, they use biochemical principles (i.e. nanopore-based single molecule sequencing~\cite{Lu2016}) that enable near real-time streaming of the raw signal as soon as the DNA molecules are ``sensed,'' usually within minutes from the process initiation. As a result, portable DNA sequencing emerges as a rapid {\it in situ} diagnostic tool, especially when DNA samples are difficult or impossible to preserve or transport. Examples include the DNA surveillance of Ebola and Zika during the recent outbreaks in Africa~\cite{Quick2016} and Brazil~\cite{Faria2016}, or successful deployments in the Arctic~\cite{Edwards2017b} and  Antarctic~\cite{Johnson2017}, in rainforests of Ecuador~\cite{Pomerantz2017}, and even on the International Space Station~\cite{Castro-Wallace2016}.

However, portable DNA sequencing and analysis are challenging for mobile systems. This is because the underlying computations, as well as data and communication intensive operations, have to be balanced to ensure the desired quality of the analysis while running in real-time, typically in the energy and bandwidth constrained environments.
%For example, XXX \todo{some good example showing our point here}.
Currently, mobile DNA sequencing is driven by the bioinformatics tools designed primarily for the desktop systems, and organized into mobile workflows in an {\it ad hoc} manner (see Section~\ref{sec:work}). While these solutions have been successful in the initial trials, they cannot be expected to scale if the underlying problems of processing speed, energy efficiency and resilience are not addressed. As the technology behind portable DNA sequencing matures and becomes more accessible, it is reasonable to assume that it will be adopted by individual consumers. Consequently, the underlying algorithms and software will have to be able to operate in the wide range of conditions, under different loads, and with varying resources, while remaining easy~to~use.

In this paper, we discuss the challenges that mobile systems and mobile computing must address to maximize the potential of portable DNA sequencing, and {\it in situ} DNA processing. Our analysis is guided by the current and envisioned applications of DNA sequencing. We look at the identified challenges from the perspective of both algorithms and systems design, and argue for careful co-design and functionality separation. We note that the paper is written from the systems perspective, for readers with no prior background in genomics or bioinformatics.

\section{Sequencing Applications}\label{sec:applications}

%In this section we discuss general applications that highlight the enabling nature of the portable DNA sequencing.

% Applications - Lou to work on this
% sequencing in hospital to sequencing on Mars, there are interesting spatial and temporal scales
% notes from meeting 8/17 goal of the paper is to introduce the potential applications, challenges and future directions
% future vision - anyone should be able to sequence anywhere anytime - take the sequencer and smartphone to a remote location and determine the microorganism communities or the presence of pathogens

%We begin with a comprehensive overview of the key applications that highlight the enabling nature of the portable DNA sequencing. We focus on two broad categories where portable DNA sequencing is crucial due to i) time constraints, and ii) spatial limitations.

We begin with a discussion of general applications that highlight the enabling nature of the portable DNA sequencing.

\textbf{Sequencing When Time is Critical:} Rapid diagnosis of infectious diseases is critical for protecting human health~\cite{Gardy2017}. The DNA sequencing can be used to identify the infectious agent, assess its responsiveness to vaccines or antibiotics, and prescribe the best treatment. In some diseases, starting the right treatment within hours is vital~\cite{Cao2016,Hewitt2017}, and when responding to epidemics, real-time genomic surveillance increases situational awareness (e.g. by tracking evolution rate and transmission patterns), helping with planning and resource allocation~\cite{Faria2016,Quick2016}. Importantly, the same principles apply in detection and mitigation of biological threats~\cite{Walter2017}. However, the current culture-based laboratory methods for identifying pathogens take days, and in fact some pathogens are difficult or impossible to grow in culture.

Because emerging portable DNA sequencers are free of these limitations, they offer significant advantage when time is critical. One excellent example is the response to the recent Ebola epidemics, where mobile laboratory based on MinIONs, transported in a standard airplane luggage, was deployed in Guinea~\cite{Quick2016}. Despite logistic difficulties, such as lack of continuous electric power and poor Internet connectivity, the laboratory became operational within two days, was generating results in less than 24 hours from receiving a sample, and provided valuable insights into disease dynamics.

\textbf{Sequencing When Location is Critical:} The samples we know the least about, and would like to study by DNA sequencing, are usually located far from established sequencing facilities. This is especially true for metagenomic studies in which communities of microbial organisms are sampled directly from their native environments, to characterize their structure and function~\cite{Quince2017}. However, many types of samples cannot be transported due to the legal and international export barriers, or other practical considerations (e.g. cost effectiveness). Furthermore, sequencing in laboratory does not allow for on-site iterative surveillance, in which sampling decisions are made in real-time. Such approaches are important when studying rapidly changing environments.

Portable DNA sequencers reached the level where they can be operated in some of the most demanding environments. For example, in one recent study, a battery-only powered laboratory, consisting of MinION  and {\it ad hoc} cluster of two laptops without Internet connection, was harnessed for {\it in situ} analysis of microorganisms found in the 100 meters deep South Wales Coalfield~\cite{Edwards2017}. Although the entire process was far from simple, the study demonstrated that DNA sequencing in remote locations is currently feasible. Other studies~\cite{Johnson2017,Pomerantz2017,Castro-Wallace2016} serve as further proof~of~principle.

\textbf{Future:} With the continuing improvements to the sequencing technology, and simplification and automation of DNA extraction and preparation protocols (see next section), we may expect that portable DNA analysis will become a ubiquitous tool. Rapid medical diagnostics, forensics, agriculture, and general exploration of microbial diversity on Earth and in outer space, are just some domains that will benefit. However, the most exciting opportunities are in consumer genomics. ONT has been promoting the idea of Internet of Living Things (IoLT)~\cite{IoLT,Waltz2017}, where anyone will be able to sequence anything anywhere, opening endless possibilities for DNA-driven discoveries. Yet, because even short DNA sequencing runs can easily deliver gigabytes of data, which may require hours to analyze, the success of IoLT will depend on the ability of mobile and cloud computing to provide~adequate~support.

%\subsection{Future}
%\todo{In the future, mobile DNA sequencing can be used (1) as an exploratory tool in remote locations on Earth and in the solar system, (2) as an early detection warning system for contaminants in air and water, and (3) as a consumer product that individuals can use to find out more about themselves, their food and their surroundings. In the long term we may expect that anyone will be able to sequence anything anywhere and this could lead to the Internet of Living Things (promoted by Oxford Nanopore Technology).}
%\begin{itemize}
%\item DNA sensors in the airplanes and other critical places (as early warning systems), DNA sequencing on spaceships and on Mars
%\item Consumer genetics, including quantified self and quantified environment around me (e.g. my food). In the long term we may expect that anyone will be able to sequence anything anywhere and this could lead to the Internet of Living Things (promoted by Oxford Nanopore Technology).
%\end{itemize}

%Infectious Disease, Water Testing, Food Safety And Efficiency, Environmental Monitoring, Supply Chain Monitoring
%Outbreak Surveillance
%Forensics
%Agriculture: Animal 
%Agriculture: Plant 
%Industrial Diagnostics
%Pharmaceuticals
%Oncology
%Reproductive Medicine
%Clinical Genetics
%Education
%Consumer Genetics (Quantified Self)
%Other

%Everyone should be able to sequence everything anywhere.

\begin{figure*}[t]
\centering
%\scalebox{0.8}{
\includegraphics[scale=0.575]{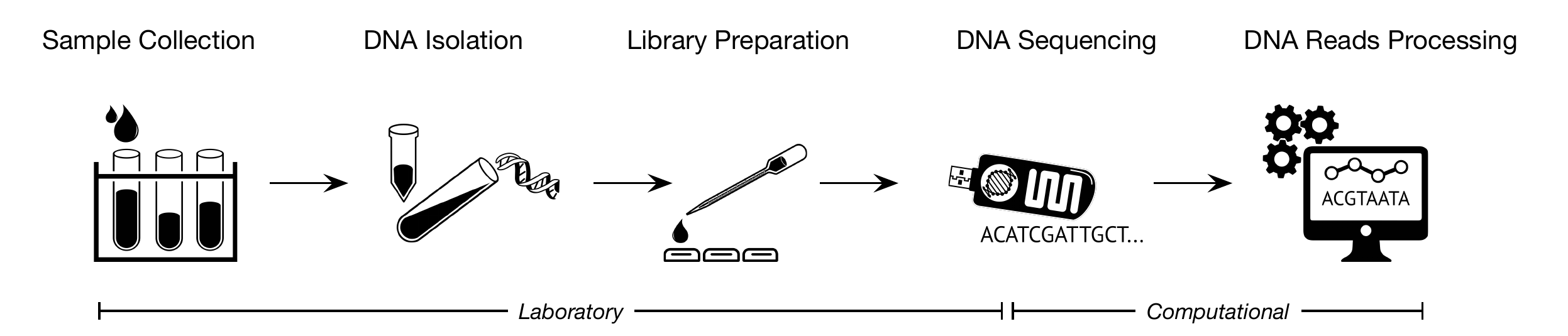}
%}
%\vspace*{-.1in}
\caption{The major steps in a typical DNA sequencing workflow.}\label{fig:workflow}
%\vspace*{-.1in}
\end{figure*}

\section{Portable vs. Benchtop}\label{sec:sequencing}

In order to understand computational challenges in portable DNA sequencing, it helps to first look at the end-to-end DNA sequencing process, and how this process differs between the current benchtop sequencing platforms and the emerging portable~sequencers.

\definecolor{Gray10}{gray}{0.9}
\begin{table*}
\centering
\footnotesize\sf
\caption{General characteristics of different steps in DNA sequencing.}\label{tab:properties}
%\vspace*{-.1in}
%\scalebox{0.8}{
\begin{tabular}{|p{0.675in}|p{0.87in}|p{0.87in}|p{0.87in}|p{0.87in}|p{0.87in}|p{0.87in}|}
\hline
\rowcolor{Gray10}
& \multicolumn{2}{c|}{\textsf{\textbf{Library Preparation}}} & \multicolumn{2}{c|}{\textsf{\textbf{DNA Sequencing}}} & \multicolumn{2}{c|}{\textsf{\textbf{DNA Reads Processing}}} \\
\rowcolor{Gray10}
& MinION & Illumina & MinION & Illumina & MinION & Illumina \\
\hline
\textsf{\textbf{Time Scale}} & Minutes-hours & Hours-days & Real-time, Minutes-hours & Days-weeks & Real-time, Minutes-hours & Batch-mode, Minutes-days \\
\hline
\textsf{\textbf{Equipment}} & \multicolumn{2}{p{1.74in}|}{Basic, portable laboratory equipment (e.g. pipettes, centrifuge)} & USB-stick-size portable device & Large benchtop machine & \multicolumn{2}{c|}{Laptop to data center} \\
\hline
\textsf{\textbf{Energy}} & \multicolumn{2}{c|}{1-2W} & 1W & Up to 10KW & \multicolumn{2}{p{1.74in}|}{No reliable analysis available,\newline sustained high load} \\
\hline
\textsf{\textbf{Software}} & \multicolumn{2}{c|}{N/A} & \multicolumn{2}{p{1.74in}|}{Proprietary firmware, drivers,\newline and control software} & \multicolumn{2}{p{1.74in}|}{Usually open source, complex string and statistical algorithms} \\
\hline
\textsf{\textbf{Advantages}} & Fast and easy\newline protocol & Protocols for very low DNA mass & Portable, real-time, long reads,\newline inexpensive device & Low cost per base, low error rate & Streaming algorithms~and~interactive sequencing & Many tested workflows available\\
\hline
\textsf{\textbf{Challenges}} & High mass\newline of input DNA & Time and labor\newline intensive & High error rate,\newline high cost per base & Short reads,\newline expensive device & High error rate & Short read length, high data volume \\
\hline
\end{tabular}
%}
\end{table*}

\subsection{How DNA is Studied}

A typical DNA sequencing workflow involves both laboratory and computational steps (see Fig.~\ref{fig:workflow} and Tab.~\ref{tab:properties}). While the specifics of the protocols executed in every step may vary, the main steps remain the same irrespective of the DNA sequencing platform.

\textbf{Sample Collection:} The first step is to obtain the material from which DNA will be extracted. The choice and quantity of material to sample, and the actual number of samples, are dictated by the particular application. Sample types include patient's blood (e.g. in epidemiology), feces (e.g. when studying gut microbial flora), water or soil (e.g. when tracking biological contaminants), etc. The samples are preserved, e.g. by freezing or adding a chemical buffer, to minimize degradation or contamination until they can be further processed. As we mentioned earlier, sometimes samples may be impossible to adequately preserve, necessitating immediate processing. 

\textbf{DNA Isolation:} In this step, the collected samples are subjected to chemical, mechanical or thermal processing to extract and purify the DNA molecules. DNA extractions performed in a laboratory with the standard commercially available kits take from minutes to hours, and in addition to the basic tools like pipettes, involve heating/cooling and specialized equipment, e.g. a centrifuge. Consequently, this step requires the access to power supply, or the use of improvised solutions.
%, for example paperfuge~\cite{Bhamla2017} instead of a desktop centrifuge. 

\textbf{Library Preparation:} The purified DNA is further processed, to make it compatible with the sensing machinery of the sequencer. This usually involves basic biochemical processing that nevertheless may require complex protocols. The step becomes even more difficult, when, to lower the cost of sequencing, instead of whole genome DNA, only specific DNA regions (e.g. corresponding to known marker genes) are to be sequenced, or if multiple samples are to be sequenced concurrently in a single run. Overall, the entire library preparation takes from several minutes to several days, and with the exception of targeted DNA sequencing, does not require additional equipment beyond the portable laboratory tools.

\textbf{DNA Sequencing:} Once the DNA library is ready, the actual sequencing can be performed. Currently, several sequencing platforms are available, for example Illumina, PacBio, Ion Torrent or Oxford Nanopore. They differ in how DNA is detected and read, which translates into differences in: sequencing speed and throughput (i.e. the number of nucleotides detected per unit of time), length of the output reads (i.e. how long DNA fragments sequencer can sense), and error rate (i.e. how many incorrectly detected nucleotides one may expect in the output). These differences are crucial, since they directly affect the downstream processing and analysis (see e.g.~\cite{Pop2009}). The current sequencers are controlled by computers, which alse receive and store output data. With the exception of the MinION, they are not portable, taking days to complete a single sequencing run in laboratory. Finally, the sequencing process involves additional consumable resources, such as biochemical reagents and flow cells -- devices in which the actual DNA sequencing happens. 

\textbf{DNA Reads Processing:} The final step is purely computational, and its goal is to first convert signal produced by a sequencer into DNA reads, and then analyze these reads for insights. Because of the volatility of DNA, and the technical limitations of the sequencing platforms, DNA is hard to sequence as a single large molecule (e.g. a chromosome). Instead, it is sequenced in fragments that the sequencer is able to sense. The raw signal produced for each detected DNA fragment is run through base calling algorithms to generate DNA reads -- the actual strings where each detected nucleotide is represented by its corresponding letter, commonly referred as base (A -- adenine, C -- cytosine, G -- guanine, T -- thymine). 

The collected DNA reads are the input to bioinformatics analysis. Here different workflows can be applied, depending on how samples had been prepared for sequencing, and what are the questions of interest. For example, {\it de novo} DNA assembly aims at reconstructing genome from input reads~\cite{Pop2009}, while metagenomic analysis uses DNA reads to detect, classify, and functionally annotate microorganisms present in the sequenced samples~\cite{Quince2017}. However, irrespective of the applied analysis, the common denominator is the reliance on compute and memory intensive string, combinatorial and statistical algorithms, ranging from massive graphs construction and traversal~\cite{Pop2009}, through clustering~\cite{Yang2011}, to large databases querying~\cite{Kim2016}. Consequently, this step requires access to non-trivial computational resources, often exceeding capabilities of a single laptop or even a~desktop~computer.

\subsection{Comparison}

To compare portable and benchtop sequencing we concentrate on the MinION, currently the only portable DNA sequencing technology, and Illumina platform, the dominating benchtop solution. We make our comparison in the context of {\it Library Preparation}, {\it DNA Sequencing}, and {\it DNA Reads Processing}, since these steps vary between platforms. Table~\ref{tab:properties} summarizes our~comparison.

% to add to the table
% DNA Isolation
% Time Scale: Hours
% Equipment: Basic laboratory equipment (pipettes, centrifuge, vortex)
% Energy: 1-2 Watts
% Software: N/A
% Challenge: environmental matrix effects DNA extraction efficiency

% Library Preparation
% Time Scale: Hours to days
% Equipment: Basic laboratory equipment
% Energy: N/A
% Software: N/A
% Challenge: Sample labeling

\textbf{Library Preparation:} The attractiveness of the MinION is in rapid sequencing protocol that can be executed in the field for genomic DNA sequencing. The protocol takes roughly 10~min, and can be automated by using portable hands-off sample preparation hardware~\cite{VOLTRAX}. However, compared to more time demanding protocols, it has limitations: it usually leads to lower quality sequencing output (i.e. with higher error rates), and it requires significant amount of input DNA ($\sim$240~ng). In comparison, protocols for the Illumina platform, while much more complex and labor intensive (fastest take $\sim$90~min, and most take hours), can be performed with an order of magnitude less DNA than MinION, without impacting sequencing quality. 

%It is difficult to directly compare different library preparation protocols, as they depend on the application and the available volume of isolated DNA. That said, 
%However, such prepared libraries may lead to lower quality sequencing output (i.e. with higher error rate) compared to more time demanding protocols, thus necessitating more intense DNA reads processing. Moreover, in the current version, the protocol requires $\sim$400~ng of input DNA. In comparison, much more complex Illumina protocols take at least $\sim$90~minutes (and most of them take hours), but can be performed with as littne as 1~ng of DNA, without impacting sequencing quality. 

\textbf{DNA Sequencing:} The MinION is based on the idea of nanopore sequencing, where DNA molecules pass through organic nanoscale sensors in a flow cell~\cite{Lu2016}. This approach has three main advantages: first, it permits portable and easy to use design with a minimal power consumption, second, it enables real-time sequencing -- the signal gathered by hundreds of nanopores in a sequencer becomes immediately available for downstream processing, third, it can produce long reads (current average is $\sim$7K bases compared to $\sim$250 bases for Illumina), and theoretically this length is limited by the physical size of DNA fragments. All this comes at the price of high error rate (anywhere between 10\%-30\% compared to $\sim$0.1\% for Illumina), which may lead to poor sequencing yield and complicates downstream analysis. Finally, because of the lower throughput, the cost of sequencing a single base is higher compared to benchtop sequencers (if we exclude the required upfront investments). This is in part because the MinION flow cells last for at most 48h of continuous sequencing. 

\begin{figure*}[t]
\centering
%\scalebox{0.8}{
\includegraphics[scale=0.625]{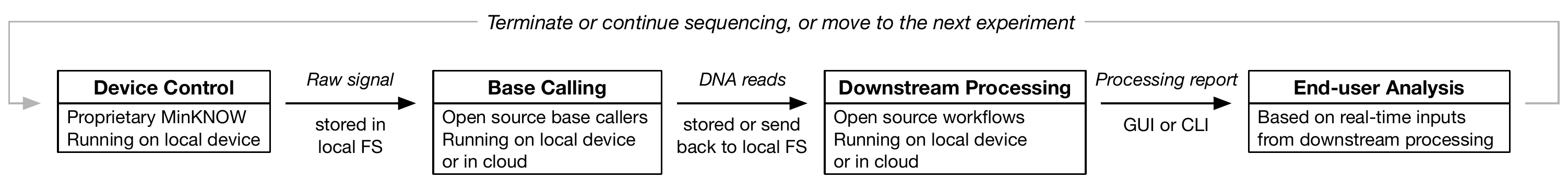}
%}
%\vspace*{-.1in}
\caption{Current MinION software workflow.}\label{fig:softflow}
%\vspace*{-.15in}
\end{figure*}

\textbf{DNA Reads Processing:} The DNA reads produced in real-time by MinION allow for flexible approach to bioinformatics analysis, with the emphasis on streaming and one-pass strategies executed outside of data centers. This makes ``interactive'' sequencing possible, where the process can be terminated as soon as sufficient DNA reads have been collected to answer questions at hand. Long reads simplify tasks such as querying of databases or DNA assembly, but the high error rate makes certain analyses (e.g. variants detection) extremely challenging, or requires more input reads to resolve uncertainties. In contrast, the Illumina platform is high-throughput and high-volume oriented. The data from a single batch run is typically processed by well established, and to some extent standardized, workflows executed in a data center close to the sequencing facility. Low error rate combined with high DNA reads volume, makes tasks such as variants detection possible. However, short reads force complex algorithms on other~tasks,~e.g.~assembly.

%However, the algorithms are computationally challenging introducing a gap into the truly real-time experience. Finally, 

%the sequencing process can be easily terminated as soon as enough 

%The long reads produced by MinION have the potential advantage of computationally easier analytics, for example, when detecting overlaps between multiple reads to recover longer DNA fragments, or when querying the DNA databases for similar annotated sequences. However, because of the higher error rate, more input reads to resolve ambiguities, especially when genetic variants such as single nucleotide polymorphisms are of interest (e.g. in medical applications).

%Moreover, certain tasks, such as 

%On the other hand Illumina provides, 
%delivered by MinION 
%require fine in certain applications, but not sufficient for variant calling. To deal with error rate long sequencing runs no dedicated for real-time and mobile processing. In contrast, Illumina has well established software base for many different applications.

\section{Mobile Computing Perspective}\label{sec:work}

We are now ready to highlight some of the open problems in portable DNA analysis as pertaining to mobile computing. To help understand the problems, we first provide overview the state of the art in mobile DNA analysis software. Then we discuss the limitations and open problems. While we base our discussion on the current MinION platform, we believe that the points we make are equally applicable to the future generation of~portable~sequencers.

\subsection{Current Software Overview}

Figure~\ref{fig:softflow} shows the general MinION software architecture. The first component is MinKNOW -- the proprietary control and signal acquisition software suite. MinKNOW is responsible for the configuration and supervision, including initiation and termination, of sequencing runs. It also receives and stores DNA signals generated by the ASIC in the sequencer's flow cell. %Currently, MinKNOW stores all digitized DNA signals using HDF5 containers on the local file system of the host device (e.g. laptop).

The second component is base calling software. Here multiple options are available~\cite{BASECALLERS}, including execution on the host device or delegation to a specialized cloud service (e.g. Metrichor~\cite{METRICHOR}). The current basecallers are very compute intensive, e.g. they typically involve Recurrent Neural Nets (RNNs), and most of the time are recommended to run in the cloud.

The final software component consists of the application specific bioinformatics tools selected by end-user. Here, bioinformatics and computational biology deliver very broad range of open source solutions to choose from, and new solutions are constantly being developed. Usually, these tools are organized into a pipeline of their own, and may involve multiple processing stages (e.g. preprocessing to remove low-quality data, querying a database to find sequences matching given DNA read, building a tree of generic relationship, etc.). Depending on the complexity, this DNA analytics may be deployed on the host machine, but more frequently it will be running in~the~cloud.

\subsection{Limitations}

The software architecture discussed above has been already used with success to showcase the promise of portable DNA sequencing. However, the approaches thus far focused on manual and {\it ad hoc} organization of the existing bioinformatics tools to perform mobile DNA analysis, without addressing many important concerns, including a systematic approach to energy, data and network management. While this is a great first step, in the long run this strategy has too many limitations to be scalable.

\textbf{Energy Management:} Energy is one of the most critical resources in mobile environments. It is especially important for mobile DNA analysis, since a DNA sequencing device is directly attached to, and draws energy from, a host device. Moreover, computational tasks executed at different stages of the sequencing workflow can run for tens of minutes to hours. However, the current software tools do not have any mechanisms to consider energy as a manageable resource. In fact, bioinformatics software tools are routinely designed under the assumption that they will be executing either on computational servers or in data centers, with abundant main memory and storage, parallel execution~capabilities, and stable power supply.

\textbf{Data Management:} DNA sequencer generates large volumes of data, which flows through various processing stages. For example, a 48h continuous run may produce up to $\sim$250~GB of output and $\sim$20~GB of temporary data, scattered across millions of files. Furthermore, if a cloud backend is used for the analysis, data needs to be transferred back and forth between a mobile device and~the~backend.

Unfortunately, data management is currently done independently by each software component. This puts an unnecessary burden on software designers -- they need to implement not only core functionality (e.g. a base calling algorithm) but also data management logic (e.g. sending data between a mobile device and a cloud, ensuring interoperability with other software, etc.). In addition, this makes it difficult to deploy new data management mechanisms rapidly, as they have to be integrated into multiple and disjoint software elements.

\textbf{Network Management:} Some of the most promising and anticipated applications of mobile DNA analysis are {\it in situ} processing scenarios, where DNA has to be completely handled at a remote location. In such scenarios, network connectivity could be sporadic and bandwidth could highly fluctuate. However, the current software is not designed to be adaptive to changing network conditions. It assumes either no network connectivity, and hence runs locally, or depends on full network connectivity, and hence assumes the always-on availability of a cloud service. Moreover, the decision to run locally or remotely is left to the end user, who must decide before executing the experiment.

\textbf{Consumables Management:} As we mentioned earlier, a flow cell is the workhorse of a sequencer. It is a consumable that can be used to analyze only a limited number of DNA samples, and within a limited time. Moreover, a flow cell degrades over time, and that translates into progressively lower sequencing throughput and potentially growing error rate. Consequently, in truly mobile setups it is necessary to manage flow cells as a scarce resource. While this problem has been recognized, currently no systematic solution exists that would offer the necessary functionality.

\subsection{Open Problems and Proposed Approach}

The majority of the limitations we identified above, are cross-cutting issues that involve multiple software components in the current workflow design. For example, all software components should have some form of data management, energy management, and network management; in order to implement a new solution that addresses a limitation for any one of these, we need to work with multiple software components and apply the solution across all of the components. This is time consuming and error prone, and thus hinders rapid innovation.

To address this challenge, we envision a new software architecture that identifies all necessary functions and separates them into different software components with clean interfaces to ensure interoperability. Specifically, we propose the architecture based on three elements: the data management layer, the DNA analytics layer, and the workflow manager. This architectural separation has the benefit of allowing different components to innovate independently from other components. It also has an advantage of simplifying software development, by allowing each component to focus on its core functionality. At the same time, it allows to accommodate the existing software, especially rich and growing set of bioinformatics tools.

\textbf{Data Management Layer:} The goal of having a separate data management layer is to free other software components from the burden of managing data on their own. Thus, the data management layer should provide all functionality related to managing DNA sequencing and analysis data. This includes 1) an interface for other components to read and store data, including the back compatibility support for the POSIX interface that the existing tools use for flat files, 2) efficient algorithms and mechanisms for data management, including discovery, monitoring and delivery, and 3) integration with cloud services for processing delegation and~data~backup.

Interesting questions arise for the design of a data management layer in all three aspects. First, for the clean slate interface design, the primary question is what kind of abstractions make the most sense for DNA sequencing and analysis. As mentioned earlier, there are mainly three types of data -- raw signals generated by a sequencer, DNA reads (strings), and analysis results (e.g. stored as data tables). Thus, perhaps the most natural interface design is to have an abstraction for each data type. Such design would allow other components to easily search, access, and when needed join data without dealing with low-level details such as file~management.

Second, once the interface is fixed, the underlying implementation can freely employ various mechanisms to manage data. For example, it is well known that DNA sequencing and analysis produce a large amount of data. It is also known that DNA has much inherent redundancy due to small alphabet (only four letters) and its repetitive nature. Thus, the data management layer can employ some compression strategies. However, it is an open question as to how best to compress this data considering the trade-offs between computation cost, computation precision and constrained storage inherent to mobile systems. The existing general strategies, as well as many DNA-specific methods that take into account DNA quality and prior knowledge of sequenced genomes~\cite{Brandon2009}, may work well, but it remains to be seen which strategy is practical in a mobile setup.

Lastly, often times DNA sequencing and analysis data needs to be shipped back and forth between a mobile device and a cloud service, for further analysis. As discussed earlier, base calling requires extensive RNNs and it is recommended to run it in the cloud. Similarly, DNA analysis is often times computationally intensive and requires much computational power and access to large reference databases. Thus, the data management layer needs algorithms and mechanisms to optimize data transfer. Here again, the existing techniques, such as similarity detection and dynamic chunking, may or may not work well.

\textbf{DNA Analytics Layer:} Ideally, the DNA analytics layer should have multiple sub-components, each implementing one algorithm relevant to DNA sequencing and analysis. This includes base calling algorithms, as well as the DNA reads processing algorithms. The goal of this layer is to allow algorithm designers to solely focus on algorithms and their implementations without worrying about other orthogonal issues, such as data or energy management.

Interesting questions arise if we consider that DNA processing can leverage both mobile devices and cloud services at the same time. This provides an opportunity to revisit current solutions and redesign them such that they become amenable to running in both domains, or in either one of the domains. In fact, we can envision a programming model to simplify implementation of such strategies. The model could provide primitives to encapsulate alternative realizations of the same DNA processing task as small migratable entities, allowing them to move across mobile and cloud domains or run in parallel if necessary. It could be further extended to account for the fact that certain DNA processing problems can be answered with different quality, trading specificity or sensitivity for computational, memory or energy performance. For instance, one of the most general questions a user may wish to ask is {\it ``what's in my pot''}~\cite{Juul2015,Kim2016}, which is to report all known organisms whose DNA has been found in a metagenomic sample, and hence sequenced. The question can be answered by classifying detected organisms at the species level (i.e. fine-grain assignment) or just at the family level (i.e. coarse-grain assignment). The fine-grain assignment will typically require large reference databases and compute intensive sequence comparison algorithms. However, the process can be accelerated by using e.g. data abbreviation techniques and clever indexing schemes at the cost of lower sensitivity and precision. By supporting such multiple task realizations of the same DNA analysis problem, the model would permit for more flexible execution paths. For example, in cases where resources are scarce, ``approximate'' tasks could provide less precise but potentially useful information, instead of waiting until resources are sufficient to deliver the detailed answer.

One additional advantage of having tasks-based analytics layer is the ability to easily deploy workflows with stream processing and speculative execution capabilities (the techniques known to improve resource utilization). We note that many of the existing algorithms, especially involving querying of reference databases, can be cast into this model with little or no effort, some other (e.g. construction of DNA assembly graphs, clustering, etc.) would require additional research and reformulation.

\textbf{Workflow Manager:} The goal of the workflow manager is to orchestrate all aspects of the DNA sequencing and analysis workflow, taking into account multiple static and dynamic factors that a user might encounter during a sequencing experiment. Many of the limitations that we discussed earlier fall into this category. Energy consumption, network connectivity, bandwidth variation, flow cell degradation, etc. all contribute to dynamically changing conditions of an experiment. Hence, the workflow manager should carefully monitor these variables and continuously make decisions on what the best course of actions is. This leads to several design questions. For example, how to monitor an experiment including not only the basic properties of a mobile device, e.g. how much energy is left or what is the current network condition, but also status of a flow cell and the quality of reads it is producing. Currently, very limited resources are available regarding flow cell monitoring, which we believe is an interesting research opportunity.

Once we have monitoring capabilities, the second question is how best to utilize available resources to get a desired outcome. This is especially important since a user conducting a DNA analysis in the field often has to make critical decisions based on limited information. For example, suppose a user is conducting DNA analysis in a remote area where there is no network connectivity. The user might wonder if she has enough energy to finish pending DNA analysis on her laptop and use the resulting data to adjust her experiment (e.g. collect more samples, etc.), or if she should move to an area where there is network connectivity to offload the analysis to a cloud and then continue experiment. The workflow manager should either assist users to make well-informed decisions, or be intelligent enough to make decisions on its own, without requiring any user intervention. In cases where DNA analysis algorithms are amenable for partitioning or migrating across different domains, the workflow manager could make fine-grained decisions of moving various computational tasks across domains directly leveraging capabilities exposed by the {\it Data Management Layer} and {\it DNA Analytics Layer}.

\section{Final Remarks}

The proposed architecture is our attempt to introduce a more systematic and scalable approach to mobile DNA analytics, by tapping into concepts known from edge computing. One potential caveat is that the proposed architecture would require reimplementation of some of the existing bioinformatics tools to fully leverage facilities in the {\it DNA Analytics Layer}. However, we believe this is feasible considering that portable sequencers are relatively new technology, with almost no algorithms designed for mobile systems. Currently, our team uses the latest release of MinION to investigate the questions we pose in this paper in the context of environmental DNA analysis.
%We hope to report on the progress in near future.

\bibliographystyle{ACM-Reference-Format}
\bibliography{hotmobile-2018}

\end{document}